\documentclass[12pt]{article}
\usepackage{amsmath}
\usepackage{times}
\usepackage{graphicx}
\usepackage{color}
\usepackage{multirow}
\usepackage[authoryear]{natbib}
\usepackage{rotating}
\usepackage{bbm}
\usepackage{latexsym}
\newcommand{\captionfonts}{\normalsize}

\makeatletter  
\long\def\@makecaption#1#2{%
  \vskip\abovecaptionskip
  \sbox\@tempboxa{{\captionfonts #1: #2}}%
  \ifdim \wd\@tempboxa >\hsize
    {\captionfonts #1: #2\par}
  \else
    \hbox to\hsize{\hfil\box\@tempboxa\hfil}%
  \fi
  \vskip\belowcaptionskip}
\makeatother   
\begin{document}
\hspace{13.9cm}1

\ \vspace{-10mm}\\

\noindent
{\LARGE Two-trace model for spike-timing-dependent \\ synaptic plasticity}

\ \\
{\bf \large Rodrigo Echeveste$^{1}$, 
Claudius Gros$^{1}$}\\
{$^{1}$ Institute for Theoretical Physics, 
Goethe University Frankfurt, Germany}\\

\thispagestyle{empty}
\markboth{}{NC instructions}
\ \vspace{-30mm}\\
%
%Abstract
%%%%%%%%%%%%%%%%%%%%%%%%%%%%%%%%%%%%%%%%%%%%%%%%%%%%%%%%%%%
\begin{center} {\bf Abstract} \end{center}
%%%%%%%%%%%%%%%%%%%%%%%%%%%%%%%%%%%%%%%%%%%%%%%%%%%%%%%%%%%

We present an effective model for timing-dependent 
synaptic plasticity (STDP) in terms of two interacting 
traces, corresponding to the fraction of activated NMDA 
receptors and the $Ca^{2+}$ concentration in the dendritic 
spine of the postsynaptic neuron. This model intends to 
bridge the worlds of existing simplistic phenomenological 
rules and highly detailed models, constituting thus a 
practical tool for the study of the interplay between 
neural activity and synaptic plasticity in extended spiking 
neural networks.

For isolated pairs of pre- and postsynaptic spikes the
standard pairwise STDP rule is reproduced, with appropriate 
parameters determining the respective weights and time 
scales for the causal and the anti-causal contributions. 
The model contains otherwise only three free parameters 
which can be adjusted to reproduce triplet nonlinearities 
in both hippocampal culture and cortical slices. We also 
investigate the transition from time-dependent to 
rate-dependent plasticity occurring for both correlated and 
uncorrelated spike patterns. 


%%%%%%%%%%%%%%%%%%%%%%%%%%%%%%%%%%%%%%%%%%%%%%%%%%%%%%%%%%%
\section{Introduction}
%%%%%%%%%%%%%%%%%%%%%%%%%%%%%%%%%%%%%%%%%%%%%%%%%%%%%%%%%%%

The fact that synaptic plasticity can depend on the precise 
timing of pre- and postsynaptic spikes \citep{Bi, Rubin}, 
indicates that time has to be coded somehow in individual neurons. 
If the concentration of a certain ion or molecule, which we 
will refer to as a {\it trace}, decays in time after a 
given event in a regular fashion, then the level of that 
trace could serve as a time coder, in the same way as the 
concentration of a radioactive isotope can be used to date 
a fossil.

A range of models have been proposed in the past that 
formulate long-term potentiation (LTP) and 
long-term depression (LTD) in terms of traces in the 
postsynaptic neurons \citep{Karmarkar, Badoual, Shouval, 
Rubin, Graupner, Uramoto}. Several of these models successfully 
reproduce a wide range of experimental results; including 
pairwise STDP, triplet and even quadruplet nonlinearities. 
Most models, however, require fitting of a large number of 
parameters individually for each experimental setup and involve 
heavily non-linear functions of the trace concentrations. 
While possibly realistic in nature, the study of 
neural systems modeled under these rules from a dynamical 
point of view becomes a highly non-trivial task. At the 
other end, the connection between predictions of 
simplified models, constructed as phenomenological rules
\citep{Badoual,Froemke}, and the biological underpinnings 
is normally hard to establish, as they usually aim to 
reproduce only the synaptic change and do away with the 
information stored in the the traces themselves.

In the present work we propose a straightforward model 
formulating synaptic potentiation and depression in terms 
of two interacting traces representing the fraction of 
activated N-methyl-D-aspartate (NMDA) receptors and the 
concentration of intracellular $Ca^{2+}$ at the postsynaptic 
spine, with the intention of bridging these two worlds. 
Having a low number of parameters and being composed 
of only polynomial differential equations, the model is 
able nonetheless to reproduce key features of LTP and LTD. 
Moreover, since the parameters of the model are easily 
related to the dynamical properties of the system, it 
permits to make a connection between the observed synaptic 
weight change and the behavior of the underlying traces.

%%%%%%%%%%%%%%%%%%%%%%%%%%%%%%%%%%%%%%%%%%%%%%%%%%%%%%%%%%%
\section{The model}\label{model}
%%%%%%%%%%%%%%%%%%%%%%%%%%%%%%%%%%%%%%%%%%%%%%%%%%%%%%%%%%%

\begin{figure}[t]
\begin{center}
 \includegraphics[width=0.6\textwidth]{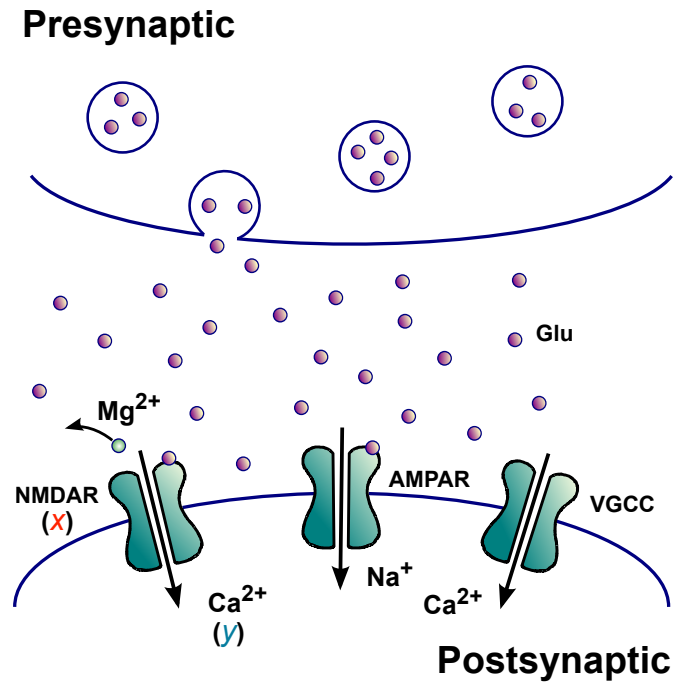}
\end{center}
\caption{Illustration of a glutamatergic synapse. Primary 
AMPA receptors (AMPAR) are directly activated by
glutamate, voltage gated calcium channels (VGCC)
by the backpropagation action potential. NMDA receptors (NMDAR) 
are also activated by glutamate and allow the influx of calcium 
if they additionally unblock, which occurs when the 
backpropagating action potential removes the blocking $Mg^{2+}$ 
ion.}
\centerline{\rule{0.8\textwidth}{0.4pt}}
\label{mechanism}
\end{figure}

Plasticity in our model will be expressed in terms of two 
interacting traces on the postsynaptic site, which we denote 
$x$ and $y$, representing the fraction of open-state 
NMDA receptors (or NMDARs) and the $Ca^{2+}$ 
concentration in the dendritic spine of the postsynaptic 
neuron, respectively. For a clarification we shortly recall 
the overall mechanism of the synaptic transmission process
in a glutamatergic synapse,
as illustrated in Fig.~\ref{mechanism}.

%---------------------------------------
%---------------------------------------

A presynaptic spike results in the release of glutamate
molecules across the synaptic cleft, which will activate
a series of receptors on the postsynaptic spine, including 
the already mentioned NMDA receptors, and 
$\alpha$-amino-3-hydroxy-5-methyl-4-isoxazolepropionic 
acid (AMPA) receptors (or AMPAR)\citep{Meldrum}. 
$Na^{+}$ ions will then flow through the AMPAR channels 
into the dendritic spine of the postsynaptic 
cell, triggering a cascade of events which may eventually 
lead to the activation of an axonal spike at the soma of the 
postsynaptic cell, and of an action potential backpropagating 
down the dendritic tree. This action potential has two
effects captured within our model: the first is
the activation of voltage-gated $Ca^{2+}$ channels (VGCC),
allowing an influx of $Ca^{2+}$ ions, resulting hence in 
an increase of the $Ca^{2+}$ concentration $y$; the second 
is the unblocking of NMDAR channels, as we detail in what 
follows.

$Ca^{2+}$ ions may flow into the postsynaptic spine also 
through the NMDAR channels \citep{Meldrum}, but for this 
to happen two conditions need to be fulfilled. NMDARs 
are activated when glutamate binds to them, which 
in turn opens the receptor's $Ca^{2+}$ permeable channel. 
The channels are said to be open when the protein's 
conformational state permits ions to flow through 
them, and closed otherwise. At resting membrane 
potential, however, $Mg^{2+}$ ions are present in the 
channel's pore, blocking the channel and preventing 
$Ca^{2+}$ ions from permeating the neuron \citep{Mayer}. 
This block is temporarily removed by a back-propagating 
action potential. For $Ca^{2+}$ to flow into the 
postsynaptic spine two conditions need hence to be fulfilled. 
The presence of glutamate in the synaptic cleft, triggered by 
a presynaptic spike, and a back-propagating action 
potential, signaling a postsynaptic spike. The NMDA 
receptors are hence the primary agents, within our model, 
for the interaction of pre- and postsynaptic neural 
activities in terms of axonal spikes. They are hence 
also the primary agents for causality within the 
STDP rule.

%%%%%%%%%%%%%%%%%%%%%%%%%%%%%%%%%%%%%%%%%%%%%%%%%%%%%%%%%%%
\subsection{Trace dynamics}\label{trace_dynamics}
%%%%%%%%%%%%%%%%%%%%%%%%%%%%%%%%%%%%%%%%%%%%%%%%%%%%%%%%%%%

We denote with $\{t_{pre}^\sigma\}$ and $\{t_{post}^\sigma\}$ 
the trains of pre- and postsynaptic spikes, respectively.
The update rules for the fraction $x$ of open but blocked 
NMDA receptors and the concentration $y$ of postsynaptic
$Ca^{2+}$ ions are then given by
\begin{equation}
\left\{
\begin{array}{lcl}
\dot x = - \frac{x}{\tau_x} + \mathit{E_x}(x)
\sum_\sigma\delta(t-t_{pre}^\sigma)  \\
\dot y = - \frac{y}{\tau_y} + (x+y_c) 
\mathit{E_y}(y)
\sum_\sigma\delta (t-t_{post}^\sigma) 
\end{array}
\right.
\label{dot_xy}
\end{equation}

where $\tau_x$ and $\tau_y$ represent the time constants 
for the decay of $x$ and $y$ respectively. In absence 
of presynaptic spikes, glutamate in the synaptic cleft
is cleared by passive diffusion and glutamate 
transporters \citep{Huang}. $Ca^{2+}$ concentration in 
the postsynaptic site will decay, in turn, in absence 
of postsynaptic spikes \citep{Carafoli}. In our model, 
each incoming presynaptic spike produces an increase 
in the number $x$ of open NMDA channels due to glutamate 
release, and the $Ca^{2+}$ concentration $y$ increases 
only when a postsynaptic spike 
is present, viz when a backpropagating action potential 
reaches the postsynaptic spine. Calcium increase in 
(\ref{dot_xy}) is composed of two terms; a constant value 
$y_c$, representing the contribution of VGCCs, and a term
proportional to the fraction of open NMDA receptors. In 
this simplified approach, every NMDAR channel still open 
from the presynaptic spike, is then unblocked by the 
backpropagating action potential. Therefore the transient 
calcium current through NMDA receptors is modeled as 
proportional to $x$. 

%---------------------------------------
\begin{figure}[t!]
\begin{center}
 \includegraphics[width=0.9\textwidth]{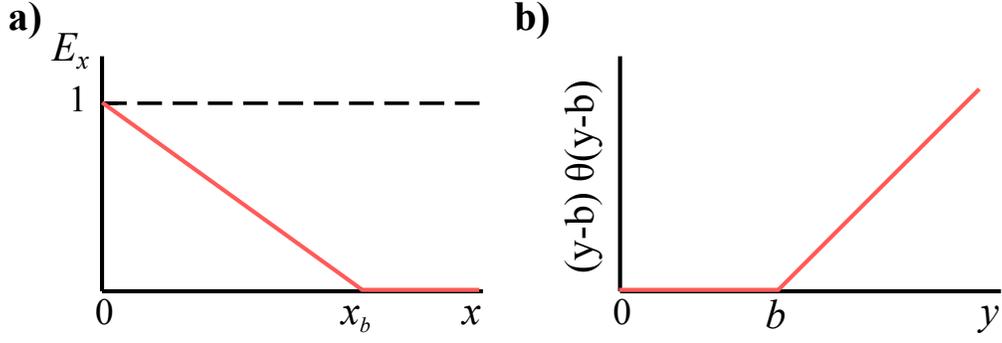}
\end{center}
\caption{\textbf{a)} Plot of the limiting factor $\mathit{E}_x$, as 
defined by (\ref{theta}), as a function of the trace concentration, 
here exemplified by $x$ (the same applies to $y$). \textbf{b)} Plot 
of the LTP threshold $(y-b)\theta(y-b)$ present in (\ref{dot_w}).}
\centerline{\rule{0.8\textwidth}{0.4pt}}
\label{limiting_factors}
\end{figure}
%---------------------------------------

\begin{figure}[t!]
\begin{center}
 \includegraphics[width=0.7\textwidth]{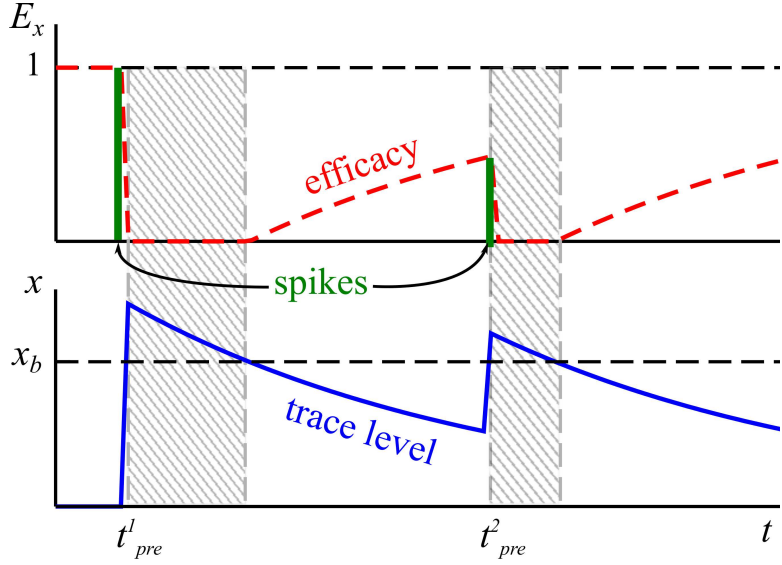}
\end{center}
\caption{Illustration of the effect of the limiting factor 
$\mathit{E}_x$ (dashed red line, upper panel),
compare Fig.~\ref{limiting_factors} and Eq.~(\ref{theta}),
on the trace dynamics (solid blue line, lower panel, compare
Eq.~(\ref{dot_xy})). Here for the $x$-trace (the behavior 
is qualitatively the same for $y$). Two spikes $t^1_{pre}$ 
and $t^2_{pre}$ are present and indicated as
solid green vertical bars, with the height being  proportional to $E_x$. 
The system ignores further incoming spike whenever $x>x_b$,
resulting in respective refractory periods (shaded grey areas).
For finite values of $x<x_b$ the efficacy of incoming spikes
is reduced proportionally to $x_b-x$.}
\centerline{\rule{0.8\textwidth}{0.4pt}}
\label{refractory}
\end{figure}

The efficacy factors $\mathit{E_x}$ and $\mathit{E_y}$ included 
in (\ref{dot_xy}) are defined as:
\begin{equation}
\mathit{E_z}(z) = \theta(z_b-z)\, \left(1-\frac{z}{z_b}\right), \qquad
\theta(z) = \left\{
\begin{array}{lcl}
0 \qquad z \leq 0 \\
1 \qquad z > 0
\end{array}
\right.
\label{theta}
\end{equation}
where $z$ is either $x$ or $y$, and determine the efficacy 
of spikes in increasing trace concentrations. For trace 
levels above the respective 
reference values $x_b$ and $y_b$ no further increase is 
possible (see Fig.~\ref{limiting_factors}\,\textbf{a)} and 
the trace concentration can only decay exponentially. This 
determines a refractory period, as shown 
in Fig.~\ref{refractory}. The duration of this period 
is in this case a function of the decay constant of the 
trace and the magnitude of the overshoot above the 
reference value. Below this level, $\mathit{E}$ will 
tend asymptotically to unity as the trace concentration decays. 
In this way, previous spikes decrease the efficacy of 
future spikes. Similar mechanisms of reduced 
spike efficacy have been proposed in the past in models 
of STDP \citep{Froemke}. 

Two forces therefore compete to drive nonlinear plasticity 
in our model: trace accumulation and spike suppression, the 
latter formulated in the present effective model via a 
saturation term. 

The update rules 
(\ref{dot_xy}) for the traces are reduced, in the 
sense that all superfluous constants have been rescaled 
away, as discussed further in the \emph{Appendix}.

%---------------------------------------
%---------------------------------------

%%%%%%%%%%%%%%%%%%%%%%%%%%%%%%%%%%%%%%%%%%%%%%%%%%%%%%%%%%%
\subsection{Update rules for the synaptic weight}
%%%%%%%%%%%%%%%%%%%%%%%%%%%%%%%%%%%%%%%%%%%%%%%%%%%%%%%%%%%

We now formulate the updating rules for the synaptic 
weight, or synaptic strength, in terms of the 
trace concentrations. To this end we consider the 
contribution of two pathways mediated by distinct enzymes 
\citep{Colbran}; which, for simplicity we will denote as 
LTP and LTD pathways. Calcium is involved in both the LTP 
and the LTD pathways \citep{Cormier, Neveu, Yang}, with 
high levels of calcium resulting in LTP and moderate and 
low levels resulting in LTD. We propose the following 
rule for the plasticity of the synaptic weight $w$,
\begin{equation}
\dot w = \alpha x (y-b) \theta(y-b) \sum_\sigma\delta(t-t_{post}^\sigma) - 
\beta xy \sum_\sigma\delta(t-t_{pre}^\sigma),
\label{dot_w}
\end{equation}
with $\theta$ being the same previously defined step function 
which, in this case, serves as a lower bound. The first term in 
(\ref{dot_w}) leads to an increase of the synaptic weight; 
it is triggered in the presence of a postsynaptic spike
and by the calcium concentration $y$, but only if $y$ 
is larger than a given threshold $b$,
see Fig.~\ref{limiting_factors}\,\textbf{b)}. A threshold 
in the calcium concentration $Ca^{2+}$ for LTP has been 
experimentally observed \citep{Cormier} and its 
dependence with the previous synaptic activity has been 
studied \citep{Huang_2}. In the present work we will consider 
a constant $b = y_c$, and we will show in the next section 
that the standard STDP curve is obtained with this choice. 

The second term in (\ref{dot_w}), in turn, leads to
decrease of the synaptic weight and needs a finite level
for both the calcium concentration $y$ and for the fraction 
of open NMDA receptors $x$ (which can be taken as a measure 
of the glutamate concentration in the synaptic cleft),
in addition to the presence of a presynaptic spike, 
which acts as a second coincidence detector as proposed by 
\cite{Karmarkar}. The parameters $\alpha>0$ and $\beta>0$ 
represent the relative strengths of these two contributions. 

The pre- and postsynaptic spikes $t_{pre/post}^\sigma$
mark the timing of the synaptic update in our model 
(\ref{dot_w}) for the synaptic plasticity. Here we 
considered $\delta$-like spikes and one needs, for 
numerical simulations using small but finite time steps, 
to update the traces via (\ref{dot_xy}) before updating 
the weights via (\ref{dot_w}).

%%%%%%%%%%%%%%%%%%%%%%%%%%%%%%%%%%%%%%%%%%%%%%%%%%%%%%%%%%%
\subsection{The pairwise STDP rule}\label{section_STDP}
%%%%%%%%%%%%%%%%%%%%%%%%%%%%%%%%%%%%%%%%%%%%%%%%%%%%%%%%%%%

In the limit of low frequencies, the traces decay to
zero in between the occurrence of two pairs of spikes,
which may hence be considered as isolated.

We denote with $\Delta t$ the time between the pre- and
the postsynaptic spike, with a positive value corresponding to
a causal pre-post order and a negative $\Delta t$ to an
anti-causal post-pre ordering. For an isolated pair of spikes 
one can easily integrate (\ref{dot_xy}) and (\ref{dot_w}), 
obtaining:
\begin{equation}
\Delta w = 
\left\{
\begin{array}{lrcl}
\alpha \mathrm{e}^{-|\Delta t|/\tau_x} 
  \left(\mathrm{e}^{-|\Delta t|/\tau_x} + y_c - b
       \right)\qquad    &\Delta t &>& 0\\
-\beta y_c \mathrm{e}^{-|\Delta t|/\tau_y} \qquad    &\Delta t &<& 0
\end{array}
\right.
\label{isolated_pair}
\end{equation}
The synaptic weight is always depressed for an anti-causal
time ordering of the spikes with $\Delta t<0$, and potentiated
for $y_c\geq b$ and a causal time ordering corresponding to
$\Delta t>0$. The LTP term becomes a simple exponential 
decay for $b=y_c$. We have chosen in our model a fully 
decoupled formulation for LTP and LTD. While the LTD term is 
always negative, the restriction on the LTP term to be always 
positive could be relaxed by removing the step function in 
equation (\ref{dot_w}). Then, with the choice $b>y_c$, a 
depression window would arise after the peak of potentiation. 
This window has indeed been observed in the past in CA1 cells 
from rat hippocampal slices \citep{Nishiyama}. By setting 
$b<y_c$, on the other hand, the decay would be composed of 
two exponentials. In the LTD term we have not included a 
threshold. Alternatively, one could replace the calcium level 
$y$ by an expression $(y-b_{LTD})\theta(y-b_{LTD})$, analogous 
to the LTP term, which is identical to the case we present for 
$b_{LTD}=0$ since $y$ is always positive. It is however worth 
discussing the cases where $b_{LTD}\neq0$. Apart from the 
step function $\theta$, the LTD threshold represents only a 
vertical shift of the negative portion of the STDP window by 
a factor $\beta b_{LTD}$. If $b_{LTD}<0$ the plot is shifted 
downwards, which means depression occurs even for isolated 
presynaptic spikes ($\Delta t\rightarrow -\infty$). This is 
usually not the case, as seen in \cite{Bi2} and \cite{Froemke}. 
If $b_{LTD}<0$, on the other hand, the plot is shifted upwards 
but, because of the step function, the $LTD$ term is always 
negative and then the tail of the exponential would be cut-off.
By looking at the experimental results in 
Figs.~\ref{STDP_theory-vs-experiment_HC} and 
\ref{STDP_theory-vs-experiment_CX}, one observes that the 
data seems in fact quite noisy to determine the exact shape 
of the decay functions. In absence of further detail, 
we have chosen in the present work to keep $b=y_c$ and no 
threshold (or a threshold at $0$) for LTD, therefore 
respecting the exponential fits proposed in the original 
papers \citep{Bi3,Froemke}.

Rewriting the constants $\alpha$, $\beta$, $\tau_x$, and
$\tau_y$ as $\alpha = A^+$, $\beta = A^-/y_c$, 
$\tau_x =2\tau_+$, and $\tau_y =\tau_-$ where 
$A^+$, $A^-$, $\tau_+$ and $\tau_-$ represent the maximal 
intensities and timescales of LTP and LTD for isolated 
spike pairs, we obtain with
\begin{equation}
\Delta w = 
\left\{
\begin{array}{lrcl}
+A^+\mathrm{e}^{-|\Delta t|/\tau_+} \qquad  &\Delta t&>&0\\
-A^-\mathrm{e}^{-|\Delta t|/\tau_-} \qquad &\Delta t&<&0
\end{array}
\right.
\label{Delta_w_isolated_pair}
\end{equation}
the classical fit for pairwise STDP proposed both in hippocampal 
and cortical neurons \citep{Bi3, Froemke}. This result is independent 
of $y_c$, $x_b$, and $y_b$ and these three parameters can be hence
be used to reproduce additional experimental observations. In what 
follows we will use the amplitudes $A^{\pm}$ as primary parameters, 
instead of $\alpha$ and $\beta$ and rewrite the plasticity
rule (\ref{dot_w}) as
\begin{equation}
\dot w = A^+ x (y-y_c)\theta(y-y_c) \sum_\sigma\delta(t-t_{post}^\sigma) 
- \frac{A^-}{y_c} xy \sum_\sigma\delta(t-t_{pre}^\sigma).
\label{dot_w_2}
\end{equation}
This is the final shape of the equation for the evolution 
of the synaptic strength that we will use throughout this work, 
it allows to interpret the results for a variety 
of spike pattern situations in terms of the known 
spike-pair STDP parameters. The representations
(\ref{dot_w_2}) and (\ref{dot_w}) are, in any case,
equivalent.

%%%%%%%%%%%%%%%%%%%%%%%%%%%%%%%%%%%%%%%%%%%%%%%%%%%%%%%%%%%
\subsection{Spike triplets}\label{sec_triplets}
%%%%%%%%%%%%%%%%%%%%%%%%%%%%%%%%%%%%%%%%%%%%%%%%%%%%%%%%%%%

The effect of a pair of pre- and postsynaptic spikes has 
been experimentally shown to depend, in a non-linear
fashion, not only on its inter-spike interval but also 
on the presence of additional spikes temporally proximal to 
the pair. The contribution of spike triplets, the simplest 
case of spike-pair interactions, cannot be described as a 
linear sum of two individual contributions of spike-pairs 
\citep{Froemke, Wang}.

In the following sections, we will study the model's results 
for either two pre- and one postsynaptic 
spikes in a pre-post-pre order, or one pre- and two 
postsynaptic spikes in a post-pre-post ordering. For 
example, with 15Post5 we denote a pre-post-pre ordering,
\begin{equation}
15\mathrm{Post}5, \qquad\quad
\left\{t_{pre}^\sigma\right\} = \{-15,5\},\qquad\quad
\left\{t_{post}^\sigma\right\} = \{0\}
\label{eq:15Post5}
\end{equation}
and with 10Pre20 a post-pre-post ordering,
\begin{equation}
10\mathrm{Pre}20, \qquad\quad
\left\{t_{pre}^\sigma\right\} = \{0\},\qquad\quad
\left\{t_{post}^\sigma\right\} = \{-10,20\}~,
\label{eq:10Pre20}
\end{equation}
where the times $t_{pre/post}^\sigma$ of
the spikes are given in milliseconds.

As for spike pairs, the weight-change induced by 
low-frequency triplets can be computed analytically, 
obtaining
\begin{equation}
\begin{array}{rcl}
 \Delta w = &+&A^+ exp\left(-\frac{\rvert\Delta t_1\rvert}{\tau_+}\right)\\
 &- &A^- exp\left(-\frac{\rvert\Delta t_2\rvert}{\tau_-}\right)
 \left[1+\frac{exp\left(-\frac{\rvert\Delta t_1\rvert}{\tau_x}\right)}{y_c}\right]
 \left[1 + exp\left(-\frac{\rvert\Delta t_1\rvert+\rvert\Delta t_2\rvert}{\tau_x}\right)\left(1 - \frac{1}{x_b}\right)\right]
\end{array}
\label{analytic_PrePostPre}
\end{equation}
for pre-post-pre triplets, and
\begin{equation}
\begin{array}{rcl}
 \Delta w = &-&A^- exp\left(-\frac{\rvert\Delta t_1\rvert}{\tau_-}\right)\\
 &+& A^+ exp\left(-\frac{\rvert\Delta t_2\rvert}{\tau_+}\right)
 \left[1 + y_c exp\left(-\frac{\rvert\Delta t_1\rvert+\rvert\Delta t_2\rvert}{\tau_y}+\frac{\rvert\Delta t_2\rvert}{\tau_x}\right)
 \left(1 - \frac{exp\left(-\rvert\Delta t_2\rvert/\tau_x\right)+y_c}{y_b}\right)\right]
\end{array}
\label{analytic_PostPrePost}
\end{equation}
for post-pre-post triplets, where we have assumed that 
the traces are below their respective reference levels,
$x_b$ and $y_b$ respetively, by the time a second spike 
arrives (the case of the second spike arriving while the 
trace is above the reference level is discussed later in 
this section). The saturation effect can reduce the effect of 
a new spike to zero but not reverse the sign, as seen 
in expression (\ref{theta}).

We see that the first term in Eqs.~(\ref{analytic_PrePostPre}) 
and (\ref{analytic_PostPrePost}), corresponding to the first pair, 
remains in both cases unchanged, by construction, with 
non-linearities appearing in the second contribution. In the second 
term of Eq.~(\ref{analytic_PrePostPre}), we find a first 
factor (the first parenthesis) corresponding to a correction 
produced by the interaction between the two traces (the 
calcium inflow through NMDAR channels), and a second factor 
corresponding to the balance between trace accumulation and 
spike suppression. In Eq.~(\ref{analytic_PostPrePost})
we also find a term balancing trace accumulation and spike 
suppression. The multiplicative factor 
$exp(\rvert\Delta t_2\rvert/\tau_x)$ inside the brackets 
comes from the way we have decided to factorize the equation, 
since $\tau_x = 2\tau_+$.

If the third spike would instead come within the respective
refractory period (see Fig.~\ref{refractory}), 
the expressions (\ref{analytic_PrePostPre}) and
(\ref{analytic_PostPrePost}) would reduce to
\begin{equation}
\begin{array}{rcl}
 \Delta w = &+&A^+ exp\left(-\frac{\rvert\Delta t_1\rvert}{\tau_+}\right)\\
 &- &A^- exp\left(-\frac{\rvert\Delta t_2\rvert}{\tau_-}\right)
 \left(1+\frac{exp\left(-\frac{\rvert\Delta t_1\rvert}{\tau_x}\right)}{y_c}\right)
 exp\left(-\frac{\rvert\Delta t_1\rvert+\rvert\Delta t_2\rvert}{\tau_x}\right)
\end{array}
\label{analytic_PrePostPre_2}
\end{equation}
for pre-post-pre triplets, and:
\begin{equation}
 \Delta w = -A^- exp\left(-\frac{\rvert\Delta t_1\rvert}{\tau_-}\right)
\label{analytic_PostPrePost_2}
\end{equation}
for post-pre-post triplets, where in Eq.~(\ref{analytic_PostPrePost_2})
the second pair is directly inhibited by the LTP threshold. 
While this last situation is not encountered for the low 
frequency triplet configurations presented in this work, 
it becomes relevant in high frequency scenarios. This 
condition could be relaxed by replacing
the strict threshold by a smooth sigmoidal.

%%%%%%%%%%%%%%%%%%%%%%%%%%%%%%%%%%%%%%%%%%%%%%%%%%%%%%%%%%%
\subsection{Interpretation of the variables and parameters in the model}\label{parameters}
%%%%%%%%%%%%%%%%%%%%%%%%%%%%%%%%%%%%%%%%%%%%%%%%%%%%%%%%%%%

The here proposed model contains a relative small number of 
variables and parameters and can be considered an effective 
approach with the biological underpinnings of STDP being 
governed by a substantially larger number of variables and 
parameters whose functional interdependences 
are naturally far more complex than the polynomial 
descriptions here proposed. Any effective model will 
however pool together within each effective variable or 
parameter several effects which might depend on a variety
of different factors in the biological neuron.

In section \ref{trace_dynamics}, we defined $x$ as the 
fraction of open but unblocked NMDAR channels. When 
paired with a postsynaptic spike, and under the 
simplifications assumed in the model, the value of 
$x$ can be then associated with a transient calcium 
current and a comparison with experimental results 
of the parameters related to $x$ would reflect this 
role. The time window for LTP, for instance, results 
in our model from the value of $\tau_x = 2\tau_+$ (as we showed in 
section \ref{section_STDP}). $\tau_x$ can then be 
interpreted in this context as the decay time of the 
transient calcium current. It has been argued by \cite{Hao} 
that the narrow window for LTP results from AMPA-EPSP in 
the postsynaptic spine. In fact, as reviewed in the 
same article, the whole spine seems to work as an electrical 
amplifier, locally prolonging the depolarization time at 
the spine. It is therefore not surprising to find different 
values of the time constants in different neurons or even 
within different synapses within the same neuron. In our 
model we do not compute AMPA currents directly and reduce the 
overall effect of the spine to the effective value of 
$\tau_x$. Similarly, $\tau_y$ represents the timescale for 
decay of the effective calcium concentration at the spine.

We have included in this work saturation terms for both 
variables $x$ and $y$. As it has been proposed in the past 
\citep{Froemke}, triplet nonlinearities in visual cortical 
neurons indicate strong suppression effects on future 
spikes by previous spikes of the train. The saturation 
terms included in the model provide one possible 
effective way of dealing with spike suppression, reducing
a biological complex phenomena further down the cascade of 
processes, leading eventually to LTP and to LTD respectively.

%%%%%%%%%%%%%%%%%%%%%%%%%%%%%%%%%%%%%%%%%%%%%%%%%%%%%%%%%%%
\section{Results for the Hippocampus}\label{Results_Hippocampus}
%%%%%%%%%%%%%%%%%%%%%%%%%%%%%%%%%%%%%%%%%%%%%%%%%%%%%%%%%%%

\begin{figure}[t]
\begin{center}
 \includegraphics[width=0.6\textwidth]{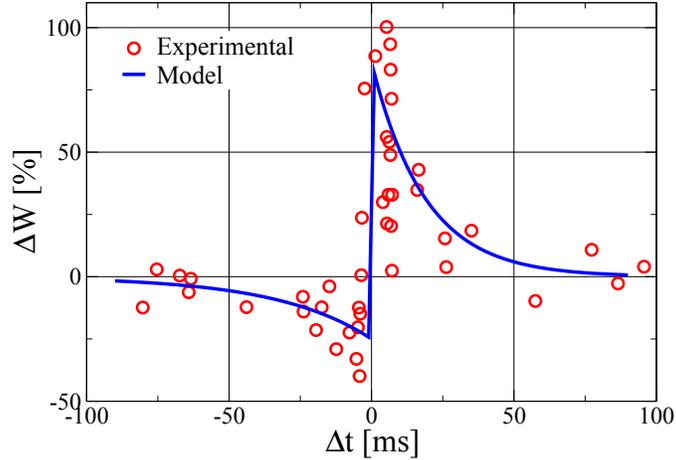}
\end{center}
\caption{Weight change after a train of 60 pairs at a constant 
frequency of $1\,\mathrm{Hz}$ as a function of the time delay 
$\Delta t$ between pre- and postsynaptic spikes. The red open 
circles are the experimental data for hippocampal neurons 
\citep{Bi2}. The continuous blue line represents the model's 
results when the parameters are set to $A^+= 0.86/60$, 
$A^- = 0.25/60$, $\tau_+=19\mathrm{ms}$, $\tau_-=34\mathrm{ms}$, 
which correspond to a fit of the experimental 
data, as presented in \cite{Bi3}.
}
\centerline{\rule{0.8\textwidth}{0.4pt}}
\label{STDP_theory-vs-experiment_HC}
\end{figure}

Our model, as defined by (\ref{dot_xy}) and
(\ref{dot_w_2}) contains overall seven adjustable 
parameters. Four of these parameters, namely
$A^+=\alpha$, $A^-=\beta/y_c$, $\tau_+$, and $\tau_-$,
enter explicitly the isolated spike-pair STDP rule 
(\ref{Delta_w_isolated_pair}) and are determined directly 
by experiment. For cultured rat hippocampal neurons 
\begin{equation}
A^+= 0.86/60,\qquad  A^- = 0.25/60,\qquad \tau_+=19\,\mathrm{ms}, \qquad 
\tau_-=34\mathrm\,{ms}
\label{parameters_A_tau}
\end{equation}
have been measured \citep{Bi3} and we will use these
experimental values throughout the hippocampus part of 
this study. In Fig.~\ref{STDP_theory-vs-experiment_HC} we 
present, as an illustration, both the experimental and the
theory results, with the latter reproducing, by construction,
the experimental fit. For the model simulation, the experimental 
protocol of 60 repetitions spaced by one second has been used. 
However, the $1\mathrm{Hz}$ frequency of spike 
pairs is so low that (\ref{Delta_w_isolated_pair}) could be 
directly used without any discernible difference.

%---------------------------------------
%---------------------------------------

Three parameters entering (\ref{dot_xy}) and (\ref{dot_w_2}), 
namely $y_c$, $x_b$, and $y_b$ are to be 
selected. In a continuous time evolution scenario, $x_b$ 
and $y_b$ determine strict maximal concentrations for 
the traces. In the discrete time scenario, overshoots are 
however possible, due to the finite increase in the traces 
after every spike. In this context, 
and in a low frequency situation, the first spike in the 
stimulation pattern is unaffected by the limiting factor, 
and only the efficacy of the following spike is reduced. 
Since $x_b$ and $y_b$ then do not affect pairwise 
STDP, they need to be selected from higher order 
contributions to the weight-change. In 
this case, we selected the values of $y_c$, $x_b$, 
and $y_b$ from triplet results, as presented in 
what follows.

\begin{figure}[t!]
\begin{center}
 \includegraphics[width=0.6\textwidth]{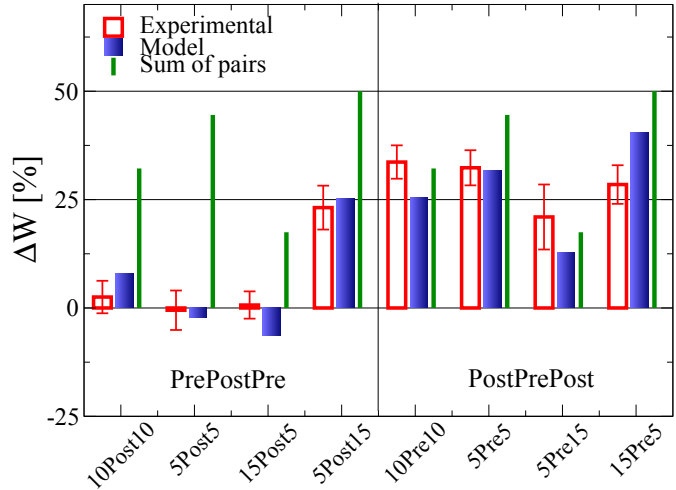}
\end{center}
\caption{Synaptic strength change in hippocampal neurons 
induced by triplets composed of either two pre- and one postsynaptic 
spike, left side of the diagram, compare (\ref{eq:15Post5}), 
or one pre- and two postsynaptic spikes, right side of the
diagram, see (\ref{eq:10Pre20}). A total
of 60 triplets are presented, with a repetition frequency
of $1\,\mathrm{Hz}$. Full blue boxes correspond to the model's 
results, empty red bars to experimental data \citep{Wang}, 
and the green lines
represent the linear addition of the PostPre and PrePost 
pairs each triplet contains via (\ref{dot_w_2}). 
Simulation parameters:
$A^+= 0.86/60$, $A^- = 0.25/60$, $\tau_+=19\mathrm{ms}$, 
$\tau_-=34\mathrm{ms}$, $y_c = 0.28$, $x_b = 0.62$, and 
$y_b = 0.66$.}
\centerline{\rule{0.8\textwidth}{0.4pt}}
\label{triplets_theory_experiment_HC}
\end{figure}

In Fig.~\ref{triplets_theory_experiment_HC} we now 
compare our results for triplets, as described in section 
\ref{sec_triplets}, with experiments for cultured rat
hippocampal neurons \citep{Wang}. The triplet stimulation 
experimental protocol consists of a regular train of 60 
triplets with a repetition frequency of $1\,\mathrm{Hz}$ 
and we use the identical protocol for the theory simulations.
We also keep the pairwise STDP parameters (\ref{parameters_A_tau})
valid for cultured rat hippocampal neurons and adjust
the remaining three free parameters $y_c$, $x_b$, and $y_b$ 
by minimizing the standard deviation (SD) between the 
numerical and the experimental results, obtaining
$y_c = 0.28$, $y_b = 0.66$, and  $x_b = 0.62$
(with an SD of $6.76$). 

%---------------------------------------
%---------------------------------------

We found that the SD varies smoothly, and relatively weakly, 
with the exact choice of the three free parameters, as can be 
expected from the analytical expressions, and that
this freedom can be used to obtain a range of
functional dependencies of the synaptic plasticity 
upon spiking frequencies, as discussed in Sect.~\ref{frequency}.

We have also included in Fig.~\ref{triplets_theory_experiment_HC} 
the expected synaptic weight changes for the case of 
a linear superposition of the two respective interspike
contributions via (\ref{dot_w_2}). One observes that the 
discrepancy between the non-linear and the linear interactions 
is much stronger for pre-post-pre than for post-pre-post 
triplets. With the former leading to an overall reduced 
synaptic weight change and the later configuration to
a substantial potentiation. It is interesting to observe here
that spike suppression, as proposed in \cite{Froemke} from 
cortical neurons cannot explain nonlinearities in hippocampus.
Suppression of the second presynaptic spike in the triplet 
would reduce depression and the overall result would be 
supralinear potentiation, contrary to the experimental 
observation. Trace accumulation is the dominant effect 
driving nonlinearities in hippocampal neurons.

%%%%%%%%%%%%%%%%%%%%%%%%%%%%%%%%%%%%%%%%%%%%%%%%%%%%%%%%%%%
\section{Results for the Cortex}\label{Results_Cortex}
%%%%%%%%%%%%%%%%%%%%%%%%%%%%%%%%%%%%%%%%%%%%%%%%%%%%%%%%%%%

\begin{figure}[t]
\begin{center}
 \includegraphics[width=0.6\textwidth]{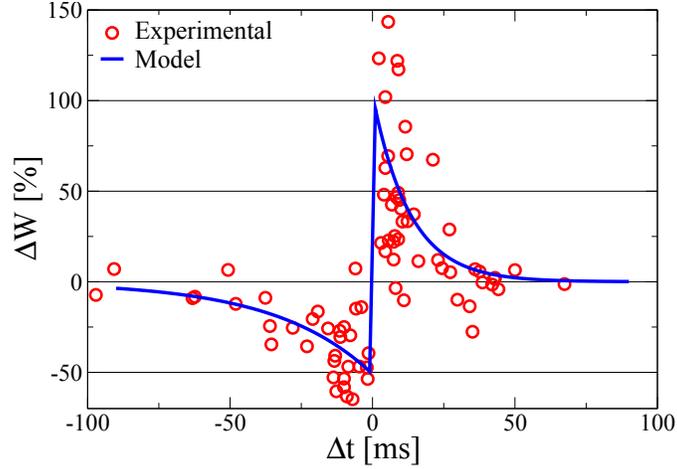}
\end{center}
\caption{As in Fig.~\ref{STDP_theory-vs-experiment_HC}, now 
for visual cortical neurons. The stimulation frequency is now 
$0.2\,\mathrm{Hz}$, as in the experiment \citep{Froemke}. 
The red open circles are the experimental data (courtesy of 
Robert C. Froemke and Yang Dan) and the continuous blue
line represents the model's results when the parameters are 
set to $A^+= 1.03/60$, $A^- = 0.51/60$, $\tau_+=13.3\,\mathrm{ms}$,
$\tau_-=34.5\mathrm\,{ms}$, corresponding to the fit of the 
experimental data presented in \cite{Froemke}.
}
\centerline{\rule{0.8\textwidth}{0.4pt}}
\label{STDP_theory-vs-experiment_CX}
\end{figure}

We repeat now the procedure presented previously for
the hippocampus, comparing the results of the proposed
plasticity rule to experimental data obtained from slices 
of the visual cortex. As in the previous section, 
the values of $A^+=\alpha$, $A^-=\beta/y_c$, $\tau_+$, and 
$\tau_-$ are determined directly by experiment. 
We use
\begin{equation}
A^+= 1.03/60,\qquad  A^- = 0.51/60,\qquad \tau_+=13.3\,\mathrm{ms}, \qquad 
\tau_-=34.5\mathrm\,{ms}
\label{parameters_A_tau_CX}
\end{equation}
as obtained by \cite{Froemke} for pyramidal 
neurons in layer 2/3 (L2/3) of rat visual cortical slices. 
Both experiment and the STDP curve
are shown in Fig.~\ref{STDP_theory-vs-experiment_CX},
where we have reproduced, for the simulation, the experimental 
protocol, using 60 repetitions at $0.2\,\mathrm{Hz}$. 
Once again, the frequency of spike pairs is so low that 
(\ref{Delta_w_isolated_pair}) could be 
directly used without any discernible difference.

%---------------------------------------
%---------------------------------------

To select $y_c$, $x_b$, and $y_b$ we once again 
resort to triplet results. In \cite{Froemke}, the change 
produced by triplets of either two pre- and one postsynaptic 
spikes or one pre- and two postsynaptic spikes was also 
measured. The data consist in this case however 
of a large set of specific triplet timing configurations,
with every individual triplet configuration measured 
once. We decided to treat all measurements on an
equal footing, fitting the complete set by minimizing
the mean square error without introducing
any further bias.

%---------------------------------------
\begin{figure}[t!]
\begin{center}
 \includegraphics[width=0.6\textwidth]{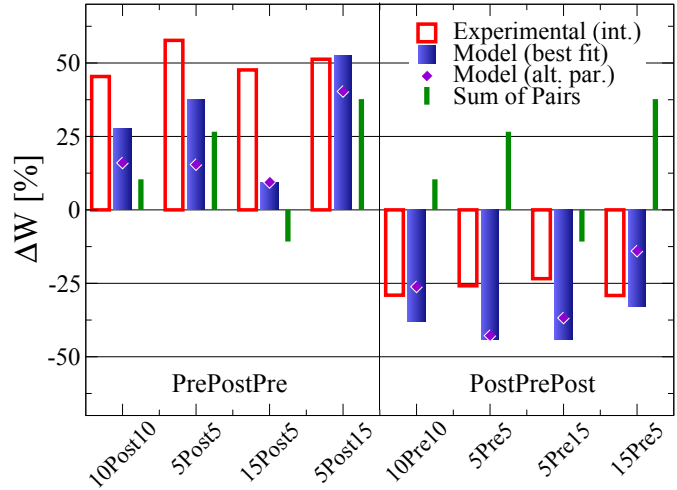}
\end{center}
\caption{As in Fig.~\ref{triplets_theory_experiment_HC}, now 
for visual cortical neurons. The stimulation frequency is now 
$0.2\,\mathrm{Hz}$ as in the experiment \citep{Froemke}.  
Full blue boxes correspond to the model's results for the best fit
of the parameters, empty red bars to experimental data,
and the green lines represent 
the linear addition of the two PostPre and PrePost pairs each 
triplet contains. With diamonds, the model's results for 
an alternative set of parameters is presented. While the 
quantitative differences are larger with this parameter choice, 
the model still qualitatively reproduces cortical triplet 
nonlinearities. Simulation parameters:
$A^+= 1.03/60$, $A^- = 0.51/60$, $\tau_+=13.3\,\mathrm{ms}$,
$\tau_-=34.5\mathrm\,{ms}$, Best fit: $y_c = 11.6$, $y_b = 10.9$, and 
$x_b = 0.5$. Diamond points: $y_c = 1.0$, $y_b = 0.9$, and 
$x_b = 0.4$.Experimental data courtesy of Robert C. 
Froemke and Yang Dan.}
\centerline{\rule{0.8\textwidth}{0.4pt}}
\label{triplets_theory_experiment_CX}
\end{figure}
%---------------------------------------

We obtain in this case $y_c = 11.6$, $y_b = 10.9$, and 
$x_b = 0.5$. The obtained SD of $37.4$ is, 
in this case, much larger than the one found for hippocampus, 
though that is partly due to the variance in the 
experimental data themselves, corresponding to individual
data points and not to averaged results. Another consequence 
of the large variance in the data is that the minimum in the 
SD is relatively broad. We will discuss these points in detail 
in what follows.

In order to compare the results for cortical neurons 
with the previous section on hippocampal neurons, as
presented in In Fig.~\ref{triplets_theory_experiment_HC},
we have performed a smooth interpolation of the set of individual
experimental results for cortical triplets by means of 
gaussian filters. In Fig.~\ref{triplets_theory_experiment_CX} 
we compare the theory results with the interpolated 
experimental data.

Contrary to hippocampal triplet results presented
in  Fig.~\ref{triplets_theory_experiment_HC}, experiments in 
cortical slices show that post-pre-post triplets lead 
to strong depression and pre-post-pre triplets to 
potentiation. Post-pre-post triplets deviate, in addition, 
somewhat more from a linear superposition of the contribution
of the two inherent spike pairs than the pre-post-pre 
configuration. While the predictions of the model presented 
in Fig.~\ref{triplets_theory_experiment_CX}) are clearly 
not as good as the ones obtained for hippocampal culture, 
they are still qualiatively in agreement with the experimental 
results, successfully capturing the asymmetry between 
post-pre-post and pre-post-pre triplets. While, there is 
still room for improvement in this regard, we 
believe it is important that the model can switch from the 
hippocampal to the cortical regime in terms of triplet 
nonlinearities.

As we previously mentioned, the data itself has a much larger 
variance in this case. To have an idea of of the variability
of the data, we computed the standard deviation of the data 
to the smooth gaussian interpolation of width 5ms that we 
used for the visual comparison of 
Fig.~\ref{triplets_theory_experiment_CX}, which 
yields an SD of $32.5$ (as compared to the SD of $37.4$ between 
model and experiment). For this reason, we believe that a 
reasonable goal in this case is to reproduce the distinct 
qualitative feature of the triplet nonlinearities, more than 
an accurate quantitative approximation. 

The optimal value of $y_c=11.6$ obtained when fitting the 
experimental triplet results, see
Fig.~\ref{triplets_theory_experiment_CX}, seems to be 
too large, in particular when compared to the one obtained 
for hippocampal neurons. This result can be traced back to
the occurrence of a broad minimum for the least-square 
fit together with a relative high variability
of the experimental data. We have hence also examined 
parameter configurations with lower values for $y_c$. Also 
included in Fig.~\ref{triplets_theory_experiment_CX} is an 
example with $y_c = 1.0$, also representing the observed 
experimental features qualitatively. We find that the 
particular cortical structure of triplets arises 
from strong saturation, being a consequence of $y_b < y_c$. 

%%%%%%%%%%%%%%%%%%%%%%%%%%%%%%%%%%%%%%%%%%%%%%%%%%%%%%%%%%%
\section{Frequency dependence}\label{frequency}
%%%%%%%%%%%%%%%%%%%%%%%%%%%%%%%%%%%%%%%%%%%%%%%%%%%%%%%%%%%

So far, we have considered only pairs or triplets of 
pre- and postsynaptic spikes coming at low frequencies and 
with very precise timings. This will not necessarily be the case 
in a natural train of spikes. It is therefore interesting to 
examine the model's prediction for spike trains with 
different degrees of correlation between pre- and postsynaptic 
spikes. A neuron usually receives input from about ten thousand 
other neurons. While the correlation of the postsynaptic neuron 
will be higher for a strong synapse driving the neuron, the 
postsynaptic neuron will in general not be correlated with all 
of its inputs. We therefore study both types of 
connections. 

We begin in section \ref{uncorrelated} by studying 
the case of uncorrelated trains of pre- and postsynaptic spikes 
and then analyze in section \ref{correlated} the case of a driving 
synapse with different degrees of correlation. In these sections 
we numerically evaluate the synaptic strength change as a 
function of the pre- and postsynaptic neuronal firing rates.

%%%%%%%%%%%%%%%%%%%%%%%%%%%%%%%%%%%%%%%%%%%%%%%%%%%%%%%%%%%
\subsection{Plasticity induced by uncorrelated spikes}\label{uncorrelated}
%%%%%%%%%%%%%%%%%%%%%%%%%%%%%%%%%%%%%%%%%%%%%%%%%%%%%%%%%%%

We begin by evaluating the synaptic change produced by 
uncorrelated trains of Poisson pre- and postsynaptic 
spikes. In these simulations we use the same parameters as 
fitted from pairwise and triplet experiments in Hippocampus 
and Cortex, refering to hippocampal and cortical 
neurons respectively. 

\begin{figure}[t]
\begin{center}
 \includegraphics[width=1.0\textwidth]{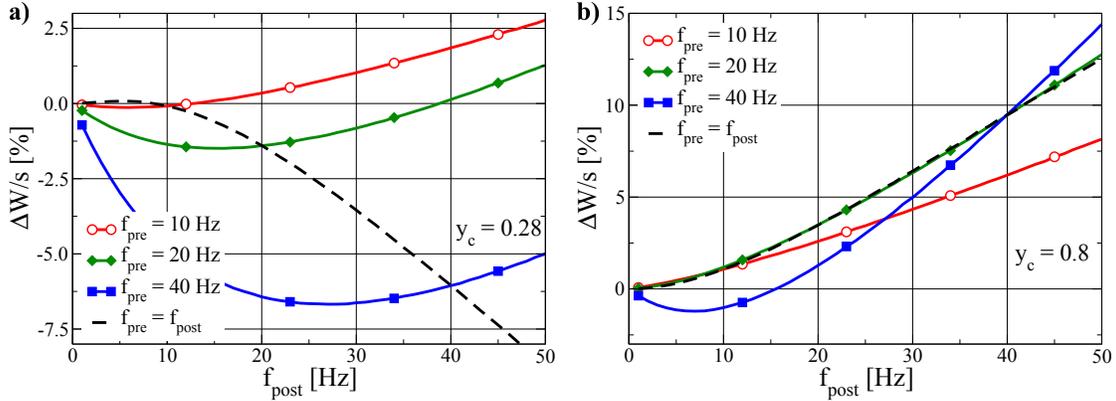}
\end{center}
\caption{Average weight change, for hippocampal neurons,
produced by one-second trains of uncorrelated 
Poisson-distributed pre- and postsynaptic spikes, 
as a function of the postsynaptic firing frequency
$f_{post}$ and for various constant presynaptic firing 
rates $f_{pre}$ (full lines).
Also included is the case for identical pre- and postsynaptic 
firing rates (dashed line). The pair-STDP values 
(\ref{parameters_A_tau}) have been used and two sets 
of values for the remaining three free parameters,
yielding both otherwise very similar results for the spike
triplets.
\textbf{a)} $y_c = 0.28$, $y_b = 0.66$, and $x_b = 0.62$.
\textbf{b)} $y_c = 0.8$, $y_b = 1.34$, and $x_b = 1.82$.
}
\centerline{\rule{0.8\textwidth}{0.4pt}}
\label{poisson_spikes_HC}
\end{figure}

The results of the simulations for hippocampal neurons 
are presented in Fig.~\ref{poisson_spikes_HC}. We 
present two kinds of plots in the diagram: a plot where 
the pre- and postsynaptic firing rates are equal, and 
plots of constant presynaptic frequency for varying 
postsynaptic firing rate. We observe in this last type, 
that the sign of the weight changes, as a function of 
the postsynaptic activity for a constant presynaptic 
frequency, generically switching from negative 
to positive at a certain threshold $\theta_H$. This 
threshold increases with rising presynaptic frequency, 
resulting in a sliding threshold. In other rate-based 
learning rules like BCM \citep{Bienenstock}, similar 
thresholds for potentiation are determined by appropriate
long term averages of the postsynaptic activity. In our 
model, $\theta_H$ is set by the level of the presynaptic 
activity, as measured on timescales of the respective 
traces. This feature would allow the neuron to adjust the 
threshold of each synapse independently, setting in each 
case the level of what constitutes a significant activity.

%---------------------------------------
%---------------------------------------

The overall synaptic change becomes Hebbian for large pre- 
and post- firing rates $f_{pre}$ and $f_{post}$, in 
the sense that it is then proportional to the product 
$f_{pre}\cdot (f_{post}-\theta_H)$. This weight change is 
influenced in a substantial way by the value of $y_c$ and 
we have presented in Fig.~\ref{poisson_spikes_HC} two sets 
of parameters, one with $y_c=0.28$ (left panel) and one 
with $y_c=0.8$ (right panel), yielding otherwise
similar SDs when fitting the experimental
triplet data ($6.76$ and $7.37$ respectively). 

Potentiation dominates for larger values of $y_c$, 
as seen in Fig.~\ref{poisson_spikes_HC}\,\textbf{b)}.
These results seem, at first sight, counterintuitive 
given the role of $y_c$ as a threshold for LTP.
Note however, that $y_c$ contributes to the increase 
in $y$ through (\ref{dot_xy}) and both LTP and LTD are 
dependent on $y$ in the plasticity rule
(\ref{dot_w_2}), with the LTD contribution being
proportional to $1/y_c$.

Comparing Figs.~\ref{poisson_spikes_HC}\,\textbf{a)}
and \textbf{b)} we observe that $y_c$ can be used to
regulate the threshold for potentiation in the rate-encoding 
limit, without changing the behavior of isolated spike 
triplets substantially. $y_c$ is hence a vehicle for
also adapting the overall postsynaptic activity
level and it would be interesting, for future 
research, to study how this regulative mechanism
would interact with other known ways to regulate 
the overall level of the postsynaptic neural activity,
such as intrinsic plasticity rules 
\citep{Triesch, Markovic, Linkerhand}.

%---------------------------------------
\begin{figure}[t]
\begin{center}
 \includegraphics[width=1.0\textwidth]{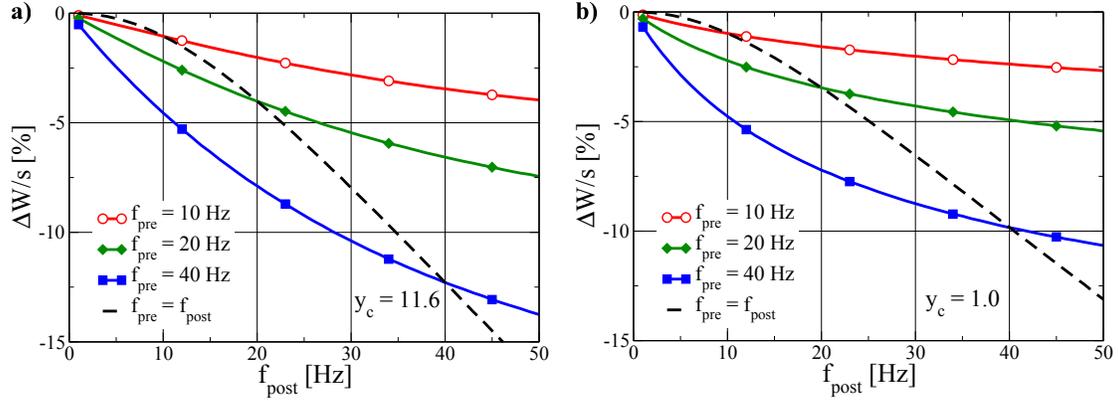}
\end{center}
\caption{As in Fig.~\ref{poisson_spikes_HC}, now 
for visual cortical neurons. Full lines show the weight change 
for specific constant presynaptic frequencies $f_{pre}$, as 
a function of the postsynaptic rate $f_{post}$. For the 
dashed line, pre- and postsynaptic firing rates are equal. 
The pair-STDP values (\ref{parameters_A_tau_CX}) have been used 
and two sets of values for the remaining three free parameters,
yielding both otherwise similar results for the spike
triplets.
\textbf{a)} $y_c = 11.6$, $y_b = 10.9$, and $x_b = 0.5$.
\textbf{b)} $y_c = 1.0$, $y_b = 0.9$, and $x_b = 0.4$.}
\centerline{\rule{0.8\textwidth}{0.4pt}}
\label{poisson_spikes_CX}
\end{figure}
%---------------------------------------

It has to be remarked that the full lines in 
Fig.~\ref{poisson_spikes_HC}, representing weight changes 
as a function of the postsynaptic frequency for a constant 
presynaptic firing rate, while of theoretical interest 
to understand the behaviour of $\theta_H$, will not 
correspond to a usual physiological functional relationship 
between the rates, at least for a driving synapse. If the 
presynaptic synapse drives the postsynaptic neuron, the 
postsynaptic activity will in general be an increasing 
function of the presynaptic rate. Here we have chosen 
$f_{pre} = f_{post}$ (the dashed lines in 
Fig.~\ref{poisson_spikes_HC} and \ref{poisson_spikes_CX})
as an illustration, but a more detailed transfer function 
should be selected for accurate and quantitative 
comparisons with experimental results. In this sense, 
the parameter configuration of 
Fig.~\ref{poisson_spikes_HC}\,\textbf{b)} shows a better 
agreement with experimental procedures, such as that of 
\cite{Sjostrom}, where potentiation is shown to become 
stronger with higher frequencies. 

No complete set of experimental results has hitherto 
been published, unfortunately, where all pairwise, triplet, 
and frequency dependent plasticity have been measured for the 
same type of synapse and with the same experimental 
stimulation procedure. A full consistency check between 
model and experiment is hence not possible to date.

In Fig.~\ref{poisson_spikes_CX} the results of numerical 
simulations for L2/3 cortical neurons for the same 
protocol of Fig.~\ref{poisson_spikes_HC} are presented. 
In this case, depression is found for all combinations of
pre- and postsynaptic frequencies, a robust prediction 
of the model. Different values of $y_c$ were selected 
to test this behavior, and in each case the rest of the 
parameters were fitted to the triplet results. In each 
case, the value of $y_b$ obtained by this fitting turned 
out to be lower than $y_c$. The $y$-trace has hence a hard 
time to overcome the threshold $y_c$ for LTP, as calcium 
increase by further spikes is prevented. As a test, if 
$y_{b}$ was artificially set to values larger than $y_c$, 
potentiation for larger frequencies was recovered but the 
fit of the experimental triplet data deteriorated 
substantially, obtaining potentiation for PostPrePost 
triplets, contrary to the experimental results. This 
indicates that triplet nonlinearities found in L2/3 
cortical neurons result from spike suppression, contrary 
to the predominant trace accumulation effect present in 
hippocampal neurons.

These results, predicted 
for L2/3 neurons as fitted from \cite{Froemke}, would 
then be in stark contrast to those 
of \cite{Sjostrom} for L5 neurons in visual cortex where 
LTP dominates for large frequencies. It should be pointed 
out, however, that the pairwise STDP plot presented in 
\cite{Sjostrom} is already different from that of 
L2/3 neurons, raising the question of to what extent 
results coming from different neurons, or obtained via 
different stimulation procedures, should be alike.

On the other hand, the prediction of overall depression 
dominating for uncorrelated spike trains in certain cortical 
neurons seems to be in line with, or at least does not 
contradict, experimental findings for deprivation experiments. 
In cortical areas, where topological maps are usually found, 
deprivation of sensory input has been shown to 
result in depression of the respective synaptic 
connections \citep{Trachtenberg, Feldman}. At the same 
time, correlation has been found to substantially decrease 
after these procedures, in areas projecting to cortex 
\citep{Linden}, suggesting that decorrelation of spike 
trains could be responsible for the observed depression in 
cortical neurons.

A possible reason behind the observed differences in these 
studies might be the stimulation protocol employed. While in 
\cite{Bi2} and \cite{Wang}, plasticity is triggered by 
eliciting the firing of the pre- and postsynaptic neurons by 
dual whole-cell clamp, in the cortical results from 
\cite{Froemke}, extracellular presynaptic stimulation is 
performed, clamping only the postsynaptic neurons. This creates 
an asymmetry between Pre-Post-Pre and Post-Pre-Post triplets.
Moreover, in the case of extracellular stimulation, the
question remains to what extent other synapses are being
affected, potentially triggering, in turn, other forms
of plasticity such as local synaptic scaling.

It is important to stress that the robust depression found 
here for higher frequencies is a direct consequence of 
the triplet results, and indeed vanishes if one uses 
hippocampal-like triplet results. The same suppression 
effect present for triplets also affects higher frequency 
trains, resulting in depression. 

For lower frequencies, the pairwise contribution dominates when 
determining the balance between potentiation and depression. 
In \cite{Izhikevich}, the authors show how a straightforward 
application of the pairwise rule to Poisson uncorrelated spike 
trains (as in our simulation), adding up linearly the effect 
of every pair in the train according to the pairwise STDP
rule with cortical parameters, always leads to depression, 
since the pairs simply sample the STDP curve which has an overall 
negative area (the opposite is true in hippocampal neurons, as 
we show below). Our model is by construction, equivalent in the 
low frequency limit to the linear pairwise model since isolated 
spikes produce no synaptic change in our model and triplets and 
higher order configurations become very infrequent if the 
frequency is low. For low pre- and postsynaptic frequencies 
the trains of Poisson spikes can be considered as pairs of 
random duration that sample the pairwise STDP curve.

As previously mentioned, the overall integrated area of the 
pairwise STDP curve for L2/3 cortical neurons is negative 
while it is positive for hippocampal neurons. One can easily 
integrate the exponentials and one obtains a relation $A$ 
for the areas:

\begin{equation}
 A = \frac{A^+ \tau_+}{A^- \tau_-}.
\end{equation}

While for hippocampal neurons $A=1.92$, we find in cortex 
$A=0.77$. This means that, in absence of higher order 
contributions (which is true if both the pre- and the 
postsynaptic frequencies are low), uncorrelated spikes will on 
average lead to depression in cortical neurons and to 
potentiation in hippocampal neurons. If the frequencies tend 
to zero, the average interspike interval will be long 
compared to the STDP window duration and the net amount of 
synaptic change, whether positive or negative, will be low.
In the following section this fact will become clear when 
the synaptic change of correlated and uncorrelated spikes 
are compared.

As the frequencies of pre- and postsynaptic spikes increase, 
the interspike period decreases and when this becomes comparable 
to the timescale of the STDP window (which is related to the 
trace timescale), the pairwise approximation will break down 
since interactions can no longer be neglected. When both the pre- 
and postsynaptic frequencies are on the order of $10\mathrm{Hz}$
the average time between a pre- and a postsynaptic spike is on 
the order of $50\mathrm{ms}$ and interactions are to be expected. 
This is where the particular models for the underlying dynamics 
will differ. Also in the work by \cite{Izhikevich}, the authors 
show with their \emph{Nearest-neighbor Implementation} that 
synaptic change goes from general depression (in the all-to-all 
implementation) to BCM-like, when they consider only the 
closest previous and posterior postsynaptic spike to each 
presynaptic spike to compute the linear sum of pairs. 

This choice, which at first glance would seem an approximation 
independent of any underlying dynamics, actually has 
strong implications for the biological underpinnings that 
could actually implement this algorithm. A first neighbor
approximation requires to hard reset any traces possibly 
present, forgetting completely anything that happened 
outside that window. The nearest-neighor implementation
has, in addition not the aim to explain triplet non-linearities
as the PrePostPre triplet protocol.

The interspike interaction is in our model driven by the 
undelying traces. We have chosen in our simulation to use 
for the frequency-dependence protocol the same parameters
obtained from pairwise and triplet fits. As observed in 
cortical PostPrePost triplets, strong suppression 
severely limits potentiation of further spikes 
(compare PostPrePost to linear superposition results). 
At high frequencies triplet interactions become relevant 
and the same suppression should then be evidenced for 
frequency dependent plasticity. We believe then, that any 
model aiming to reproduce time-dependent plasticity up to 
triplet order as measured by \cite{Froemke}, should show 
depression also for high frequencies in cortical neurons.

%%%%%%%%%%%%%%%%%%%%%%%%%%%%%%%%%%%%%%%%%%%%%%%%%%%%%%%%%%%
\subsection{Plasticity induced by correlated spikes}\label{correlated}
%%%%%%%%%%%%%%%%%%%%%%%%%%%%%%%%%%%%%%%%%%%%%%%%%%%%%%%%%%%

So far we have analyzed the effect of 
uncorrelated spikes on the synaptic weight change both 
in hippocampal and in cortical neurons 
(Figs.~\ref{poisson_spikes_HC} and \ref{poisson_spikes_CX}).
It is however interesting to see the predictions of the model 
for a strong synapse driving the postsynaptic neuron. In this 
case pre- and postsynaptic spikes should be correlated,
at least partially, together with a certain positive delay. 

To reproduce this effect with our model, we 
simulated trains of correlated spikes where, with each 
presynaptic spike, a postsynaptic spike can occur with 
probability $p$, after a certain delay $d$. As an example,
if every presynaptic spike triggers a postsynaptic spike 
then $p = 1$. This would mean, however that the postsynaptic 
frequency $f_{post}$ changes with $p$ ($f_{post} = p$ $f_{pre}$). 
In order to compare our results for different values of $p$ 
and keep $f_{post}$ independent of $p$, we complete the 
train of postsynaptic 
spikes with Poisson spikes of frequency $(1-p)f_{pre}$. 
In this way, $f_{post}$ is independent of $p$, which now 
regulates the degree of correlation between pre- and 
postsynaptic spikes: $p=0$ represents the fully decorrelated 
case, since all the postsynaptic spikes are drawn from 
the poisson distribution, and $p=1$ represents the fully 
correlated case already mentioned.

\begin{figure}[t!]
\begin{center}
 \includegraphics[width=1.0\textwidth]{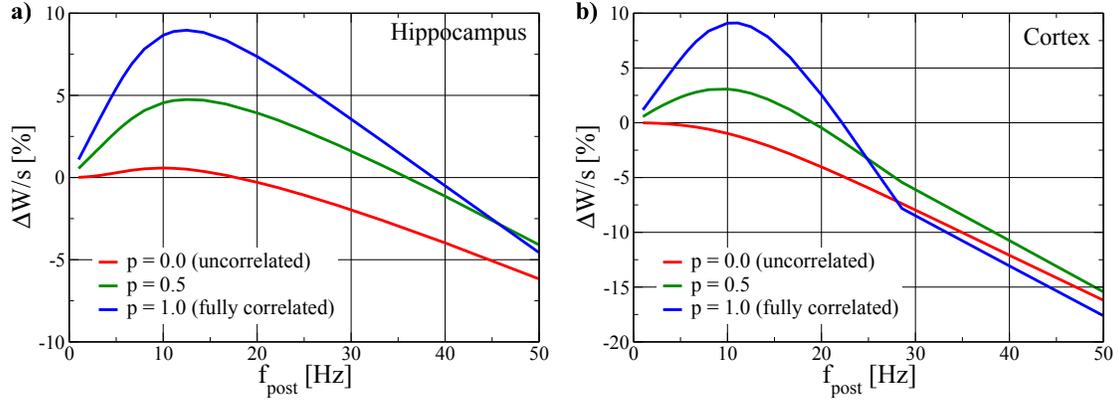}
\end{center}
\caption{Simulation results of the synaptic weight change 
induced by a one-second train of spikes with different 
degrees of correlation, as a function of the frequency 
$f = f_{pre} = f_{post}$. The fraction $p$ of correlated 
spikes takes the values $0$, $0.5$, and $1$ in these 
plots. The delay between pre- and postsynaptic spikes was 
taken to be $5\mathrm{ms}$ in these simulations. The case 
$p=0$ corresponds to the dashed lines in 
Figs.~\ref{poisson_spikes_HC} and \ref{poisson_spikes_CX}.
\textbf{a)} The pairwise hippocampal parameters 
(\ref{parameters_A_tau}) were used, together with 
$y_c = 0.28$, $y_b = 0.66$, and $x_b = 0.62$,
corresponding to the triplet fit.
\textbf{b)} Here the pairwise cortical parameters 
(\ref{parameters_A_tau_CX}) were used, together with 
$y_c = 11.6$, $y_b = 10.9$, and $x_b = 0.5$,
corresponding to the triplet fit.}
\centerline{\rule{0.8\textwidth}{0.4pt}}
\label{correlated_spikes}
\end{figure}

In Fig.~\ref{correlated_spikes} we present the synaptic 
changes produced by trains spikes for different values of 
$p$. In this case, results for a delay of $5\mathrm{ms}$  
are presented. The same tests were performed with delays from 
$2\mathrm{ms}$ to $10\mathrm{ms}$ with only quantitative but 
not qualitative differences.

%---------------------------------------
%---------------------------------------

We observe in Fig.~\ref{correlated_spikes}, both for 
hippocampal and cortical neurons, that correlated spike 
trains induce an increasing amount of potentiation for low 
to intermediate frequencies ($\sim1 -10\mathrm{Hz}$). In 
the correlated scenario, and since in this case we are 
simulating a driving synapse, postsynaptic spikes follow 
presynaptic spikes in a causal order. When the frequency 
is higher than $10\mathrm{Hz}$, the interspike period 
becomes comparable to the STDP time window and each 
postsynaptic spike will also ''see'' the following 
presynaptic spike, thus triggering the LTD term. 
Depending on the trace saturation constants, LTD or 
LTP will eventually dominate for large frequencies. 
If LTD dominates, depression results and after a 
certain reversal frequency the behavior is switched from 
Hebbian to Anti-Hebbian. This is the case for the triplet 
fitted values presented in Fig.~\ref{correlated_spikes}. 

It is important to note that the model is also able to 
produce Hebbian behavior within the entire physiological range 
of activities by changing $y_c$. The smaller the saturation 
effects are, the larger this reversal frequency becomes. In fact, 
with the second set of parameters used in 
Fig.~\ref{poisson_spikes_HC}\,\textbf{b)}, no such reversal is 
found within physiological frequencies (not shown here). 

If observed, such a reversal, which is yet another side of the 
suppression effect, would have the benefit of being self-stabilizing, 
tuning synaptic strength to help keep neural activities bound.

The kink observable for the fully correlated curve 
($p = 1.0$) of \ref{correlated_spikes}~\textbf{b)} 
results from the particularly strong suppression effects 
in cortical neurons, captured in our model by $y_b < y_c$. 
For frequencies above a certain threshold 
($\sim28\mathrm{Hz}$) the trace concentration $y$ is 
always below the $y_c$ and LTP never triggers. 

A fundamental difference between plots \textbf{a)} and 
\textbf{b)} is the different qualitative behavior between 
correlated and uncorrelated spikes. While in hippocampal 
neurons, Hebbian, increasing potentiation, is always 
present for low to intermediate frequencies (whether the 
spikes are correlated or not), in cortical neurons, 
our model predicts uncorrelated spikes always produce 
depression and therefor Hebbian learning requires the 
neurons to be at least partially correlated.

%%%%%%%%%%%%%%%%%%%%%%%%%%%%%%%%%%%%%%%%%%%%%%%%%%%%%%%%%%%
\section{Comparison to other models}
%%%%%%%%%%%%%%%%%%%%%%%%%%%%%%%%%%%%%%%%%%%%%%%%%%%%%%%%%%%

The problem of formulating plasticity in terms of the 
specific timing of pre- and postsynaptic spikes can be 
approached at different levels of detail and accuracy, 
ranging from simplistic phenomenological rules to detailed 
and complex models describing the different steps of the 
biological machinery responsible for STDP. In sections 
\ref{Results_Hippocampus} and \ref{Results_Cortex} 
the comparison of our model to simple forms of 
phenomenological rules has already been established, 
noting that linear combinations of spike pairs are 
generically not sufficient to explain the experimentally 
observed triplet non-linearities. 

We have also shown that, while linear combinations of 
pairs, plus additional suppression, is enough to explain 
the triplet nonlinearities of cortical neurons, as shown in 
\cite{Froemke}, hippocampal triplet non-linearities cannot 
be explained by suppression and a trace accumulation 
mechanism seems to be taking place in these synapses. In 
any case, this kind of phenomenological rules are not 
likely to generalize well to arbitrary spike patterns 
since no information of the underlying plasticity 
mechanism is present in the formulation.

Other models, like \cite{Albers} and \cite{Pfister}, present 
interesting dynamical formulations of plasticity in 
terms of generic decaying markers or traces, but do 
not attempt to establish a link to the biological 
underpinnings of STDP. Calcium concentration and NMDA 
receptors have been shown to play a central role in 
time-dependent LTP and LTD, and we therefore believe it is 
important to formulate plasticity in those terms. Our 
model, though simplified, is formulated in terms of these 
key ingredients and may therefore help to bridge the worlds 
of functional and realistic models.

An alternative approach has been proposed by
\cite{Appleby}, where plasticity is described 
in an ensemble-based formulation. The authors there argue 
that the observed synaptic changes produced by standard 
protocols cannot be explained at a single synapse level, 
but rather state that the observed results arise at a 
population level. The authors then show how the pairwise 
STDP curve can be recovered at the ensemble level from 
all or nothing potentiation or depression at the single 
synapse level. The dependency of the model to the specific 
timing of triplets is however in this case not computed.

The model we present in this work belongs to the family 
of calcium-based spiking-neuron models. Within this 
family, models formulating synaptic plasticity exclusively 
in terms of the calcium levels \citep{Uramoto,Graupner}, 
while tuned to reproduce a variety of 
experimental results, tend to show paradoxical results when 
tested in other setups. The model presented by \cite{Uramoto}, 
for example, predicts synaptic changes even when only 
postsynaptic spikes are present. The model in 
\cite{Graupner}, in turn, shows plasticity also when  
either pre- or postsynaptic spikes are absent, 
since both pre- and postsynaptic spikes contribute directly 
to the calcium level in this model, without the need of 
coincidence. To avoid this, in our model we demand the 
simultaneous presence of both pre- and postsynaptic spikes 
for plasticity to arise, being proportional to the products 
of traces $x$ and $y$ in our rule. We believe this to be an 
important feature for simulations in situations of complex 
spike patterns where the pre- and postsynaptic firing rates 
do not necessarily match.

%%%%%%%%%%%%%%%%%%%%%%%%%%%%%%%%%%%%%%%%%%%%%%%%%%%%%%%%%%%
\section{Discussion}
%%%%%%%%%%%%%%%%%%%%%%%%%%%%%%%%%%%%%%%%%%%%%%%%%%%%%%%%%%%

We propose a basic trace model for timing-dependent 
plasticity that incorporates, in a first order 
approximation, the fundamental mechanisms acknowledged to 
be taking place in STDP. We show that the model 
successfully captures several main features of time-dependent 
plasticity, including the standard shape for low frequency 
pairing, experimentally observed triplet nonlinearities, 
and large frequency effects.

The decay constants for the two traces and the relative 
intensities of LTP and LTD can be extracted directly from 
the standard STDP curves, as measured for isolated pairs. 
The model is left hereafter with only three further 
parameters, which can be used to fit higher order 
contributions to plasticity. The model successfully 
reproduces the distinct and contrasting nonlinearities 
found in both hippocampal cultures and cortical 
slices.

While the model predicts a similar frequency dependence 
for correlated (or partially correlated) pre- and postsynaptic 
spikes both for hippocampal and cortical neurons, the effect 
of uncorrelated spikes (although smaller) differs 
qualitatively in these two types of neurons. In this case, 
the sign of the resulting plasticity depends for lower 
frequencies on the overall area of the pairwise STDP curve, 
resulting in potentiation for hippocampal neurons and 
depression in cortical ones, and for higher frequencies 
on the balance between spike suppression and trace accumulation.

We show that the model is able to reproduce typical 
frequency-dependencies for uncorrelated spikes, while 
fitting pairwise and triplet hippocampal parameters. 
We do also find, that fitting triplet results for 
L2/3 cortical neurons invariably leads to depression,
for higher frequencies and of uncorrelated spikes, contrary 
to observations in L5 neurons. The question then arises, 
to which extent plasticity results for different neurons,
or performed under different stimulation conditions, can 
be expected to match. It seems then essential to have available 
a consistent sets of experiments where pairwise, triplet, and 
frequency results are measured for the same type of neuron and 
with the same stimulation protocol. Otherwise one runs the 
risk of possibly trying to build a complete picture out of 
mismatching parts. In this sense, we hope that our predictions 
serve as a motivation to revisit and to complete triplet and 
frequency dependent studies for different types of neurons.

In order to compare our results with rate encoding plasticity 
models, we have also shown, by setting the presynaptic frequency 
to a constant value, that the amount of synaptic change is 
proportional, for hippocampal neurons, to the product of the 
activities, with a threshold that depends on the presynaptic 
firing rate. While the system lacks a longer term average 
threshold, as the one present in BCM, the presynaptic activity 
acts as a value of reference for the significant level of activity. 
If the postsynaptic activity exceeds this level then 
potentiation occurs, otherwise, depression arises.

For correlated spikes, we have shown that the model leads to
similar results for hippocampal and cortical neurons, with 
an initial hebbian behavior for small to medium frequencies 
and, depending on choice for the parameters, a reversal to 
anti-hebbian behavior for large frequencies, which could have 
the virtue of being self-limiting, avoiding runaway growth 
of synaptic connections. It has been shown recently
\citep{Echeveste}, that this self-limitation results naturally, 
for rate encoding neurons, from the stationarity principle for 
Hebbian learning. By tuning the value of $y_c$, the reversal 
frequency can be made larger, to the point of producing Hebbian 
behavior within the entire physiological range of activities.

The simplicity of the here proposed model makes it a useful 
tool for simulations and studies of the dynamical 
properties of networks adapted via these rules. Firstly, the 
calculations required are straightforward, making it 
computing time efficient. The relative small number of free 
parameters and their direct link to both the biophysical 
properties of the postsynaptic complex and to the dynamical 
features of the trace dynamics makes it suitable when 
studying the interplay between neural and synaptic dynamics 
in  neural systems.

%%%%%%%%%%%%%%%%%%%%%%%%%%%%%%%%%%%%%%%%%%%%%%%%%%%%%%%%%%%
\section*{Acknowledgments}
%%%%%%%%%%%%%%%%%%%%%%%%%%%%%%%%%%%%%%%%%%%%%%%%%%%%%%%%%%%

The authors acknowledge Robert C. Froemke and Yang Dan for 
the experimental data from cortical visual neurons.

%%%%%%%%%%%%%%%%%%%%%%%%%%%%%%%%%%%%%%%%%%%%%%%%%%%%%%%%%%%
\section*{Appendix: Dimensionality Analysis}
%%%%%%%%%%%%%%%%%%%%%%%%%%%%%%%%%%%%%%%%%%%%%%%%%%%%%%%%%%%

In section \ref{model} we could have started by initially 
denoting by $x'$ and $y'$ the fraction of NMDA receptors and 
the $Ca^{2+}$ concentration, respectively, where the time 
evolution of these traces is written as:

%---------------------------------------
\begin{equation}
\left\{
\begin{array}{lcl}
\dot x' = - \frac{x'}{\tau_x} + c_1  \mathit{E_x}
\sum_\sigma\delta (t-t_{pre}^\sigma) \\
\dot y' = - \frac{y'}{\tau_y} + (c_2 x'+y'_c) \mathit{E_y}
\sum_\sigma\delta (t-t_{post}^\sigma)
\end{array}
\right.
\label{dot_xy_2}
\end{equation}
%---------------------------------------

where $\tau_y$ and $\tau_x$ represent the time constants in 
the decay of $x'$ and $y'$, and now two additional 
parameters $c_1$ and $c_2$ are included. $c_1$ represents 
the increase in $x'$ caused by a single presynaptic spike 
(which in this simplified model we assume constant) and 
$c_2$ represents the increase, per unit of $x'$, in $y'$ 
concentration. $y'_c$ is the constant contribution per 
postsynaptic spike to $y'$ of the voltage-gated $Ca^{2+}$ 
channels. Again a spike efficacy $\mathit{E}$ is included that 
limits trace concentrations, where $\mathit{E}$ is still 
calculated as in (\ref{theta}).

Now, by a change of variables:

%---------------------------------------
\begin{equation}
x = x'/c_1, \qquad
x_b = x'_b/c_1, \qquad
y = y'/c_1c_2, \qquad
y_c = y'_c/c_1c_2, \qquad
y_b = y'_b/c_1c_2
\label{change}
\end{equation}
%---------------------------------------

we re-write (\ref{dot_xy_2}) as:

%---------------------------------------
\begin{equation*}
\left\{
\begin{array}{lcl}
\dot x = - \frac{x}{\tau_x} + \mathit{E_x}
\sum_\sigma\delta (t-t_{pre}^\sigma) \\
\dot y = - \frac{y}{\tau_y} + (x+y_c) \mathit{E_y}
\sum_\sigma\delta (t-t_{post}^\sigma) 
\end{array}
\right.
\end{equation*}
%---------------------------------------

which is the exactly the expression presented in section 
\ref{model}. By this procedure we have reduced the number 
of parameters for the trace evolution to 5.

%%%%%%%%%%%%%%%%%%%%%%%%%%%%%%%%%%%%%%%%%%%%%%%%%%%%%%%%%%%

\end{document}